# Neutron energy spectrum from 120 GeV protons on a thick copper target

Nobuhiro Shigyo[1], Toshiya. Sanami[2], Tsuyoshi Kajimoto[1], Yosuke Iwamoto[3]
Masayuki Hagiwara[2], Kiwamu Saito[2], Kenji Ishibashi[1], Hiroshi Nakashima[3], Yukio Sakamoto[3]
Hee-Seock Lee[4], Erik Ramberg[5], Aria A. Meyhoefer[5], Rick Coleman5, Doug. Jensen[5]
Anthony F. Leveling[5], David J. Boehnlein[5], Kamran Vaziri[5], Nikolai V. Mokhov[5]

[1] Kyushu University, 744 Motooka, Fukuoka 819-0395 Japan
[2] High Energy Accelerator Research Organization, 1-1 Oho, Tsukuba 305-0801 Japan
[3] Japan Atomic Energy Agency, 1-1 Shirakata Shirane, Tokai-mura, Ibaraki 319-0395 Japan
[4] Pohang Accelerator Laboratory, Pohang, Kyungbuk 790-784, Korea
[5] Fermi National Accelerator Laboratory, Batavia, IL 60510-5011 USA

**Abstract**

*Neutron energy spectrum from 120 GeV protons on a thick copper target was measured at the Meson Test Beam Facility (MTBF) at Fermi National Accelerator Laboratory. The data allows for evaluation of neutron production process implemented in theoretical simulation codes. It also helps exploring the reasons for some disagreement between calculation results and shielding benchmark data taken at high energy accelerator facilities, since it is evaluated separately from neutron transport. The experiment was carried out using a 120 GeV proton beam of 3E5 protons/spill. Since the spill duration was 4 seconds, proton-induced events were counted pulse by pulse. The intensity was maintained using diffusers and collimators installed in the beam line to MTBF. The protons hit a copper block target the size of which is 5cm x 5cm x 60 cm long. The neutrons produced in the target were measured using NE213 liquid scintillator detectors, placed about 5.5 m away from the target at 30˚ and 5 m 90˚ with respect to the proton beam axis. The neutron energy was determined by time-of-flight technique using timing difference between the NE213 and a plastic scintillator located just before the target. Neutron detection efficiency of NE213 was determined on basis of experimental data from the high energy neutron beam line at Los Alamos National Laboratory. The neutron spectrum was compared with the results of multi-particle transport codes to validate the implemented theoretical models. The apparatus would be applied to future measurements to obtain a systematic data set for secondary particle production on various target materials.*

**Introduction**

Behavior of high energy radiation above several tens GeV region is interesting to study high energy accelerator shielding, space technology and so on. The Japanese-American Study of Muon Interactions and Neutron detection (JASMIN) collaboration [1] has been acquired shielding data and evaluated neutron production process implemented in theoretical simulation codes. In a recent result of the collaboration, the experimental neutron energy spectra at the transmission experiment for 120 GeV proton beam consists of components below several GeV. The MARS code [2], which is one of multi-particle transport code, reproduces the trend of the experimental data [3]. Neutron (and proton) production mechanism by high energy radiation as a source term is essential in order to understand behavior of high energy radiation in shielding materials.

Neutron production double-differential cross sections and thick target yields have been measured below several tens GeV [4-6]. However, there is no experimental data above several tens GeV. It is important to obtain double-differential thick target neutron and proton yields experimentally to evaluate the theoretical models in the multi-particle transport codes. In this paper, a feasibility study of experiment on double-differential neutron thick target yields by 120 GeV proton incidence was described. The preliminary experimental data was compared with the calculation results.

**Experiment**

The experiment was carried out at the MT6-2 area of Fermilab Test Beam Facility (FTBF), Fermi National Accelerator Laboratory. The schematic view of the experimental area and the detector arrangement are illustrated in Figure 1.

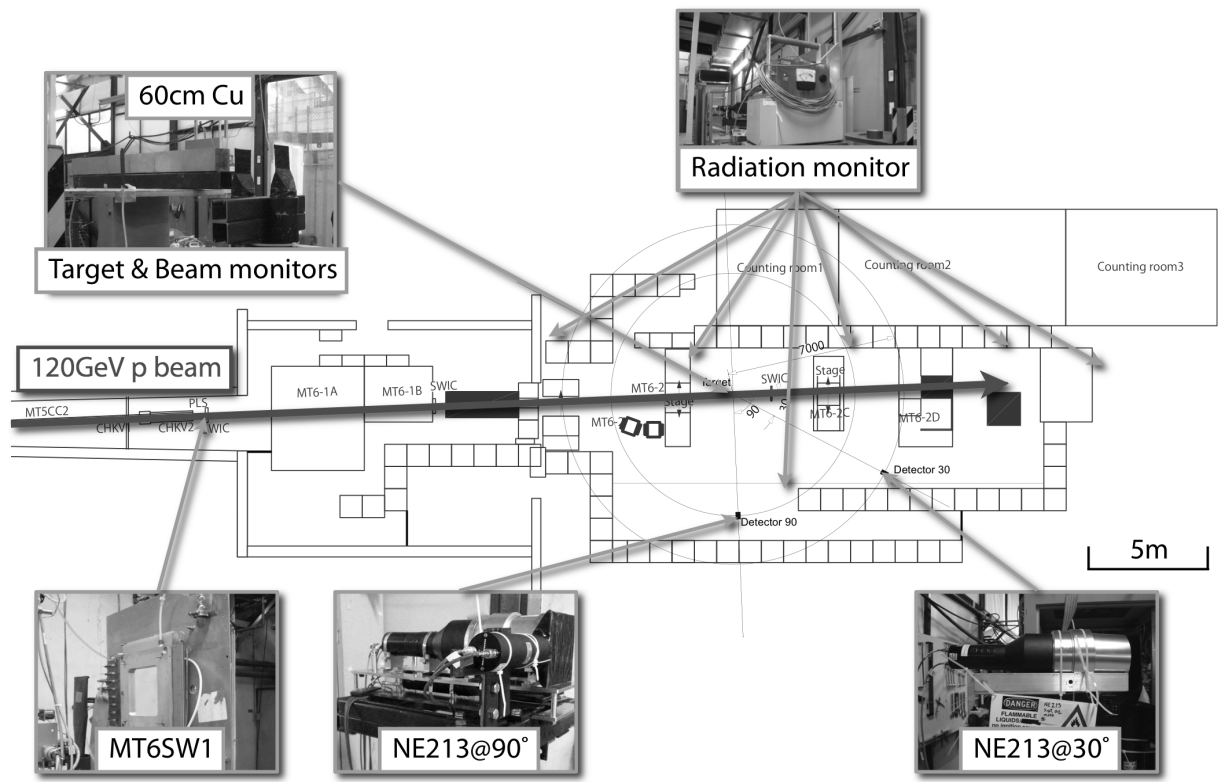

Figure 1: Experimental Setup at MT6-2 Area in Fermilab Test Beam Facility (FTBF).

A copper block which size was 5 cm x 5 cm x 60 cm long was chosen as a target. Measurement with target-in and target-out were performed to eliminate background neutron events.

Two NE213 liquid organic scintillators (12.7 cm thick and 12.7 cm long) were adopted as a neutron detector and placed at 30° and 90°. A plastic scintillator was set in front of each NE213 scintillator to distinguish neutron events with charged particle ones. Neutron energy was determined by the time-of-flight method. The flight paths were 5.5 m at 30°and 5 m at 90°, respectively. Some radiation monitors were installed to observe radiation level in the experimental area to keep the safety limit.

The 120 GeV incident proton beam was monitored by multiwire proportional counters (MT6SW1) on the upstream on the beam line to see the beam profile and three plastic scintillators in front of the target to count the number of incident proton. The average of the number of proton was limited to $3 \times 10^5$ particles/spill because of the radiation safety limitation.

Two types of data acquisition system (DAQ) were used in the measurement. One was an ordinary CAMAC DAQ to collect the integral light output from scintillator. The other one was a wave form digitizer DAQ which gets wave form of light output from scintillator as a function of time to increase actual counting rate. Both DAQ system acquired the neutron flight time between the target and the scintillator.

The time structure of proton beam consists of three levels as shown in Figure 2. The smallest structure was the bunch of 19 ns interval. The Second one was called the train which included 20 – 36 bunches and had 11 μs interval. The largest structure was the spill which contained trains of 4 seconds. This meant that the beam was available in just 4 seconds a minute.

**Figure 2: Time structure of proton beam at MT6-2 area.**

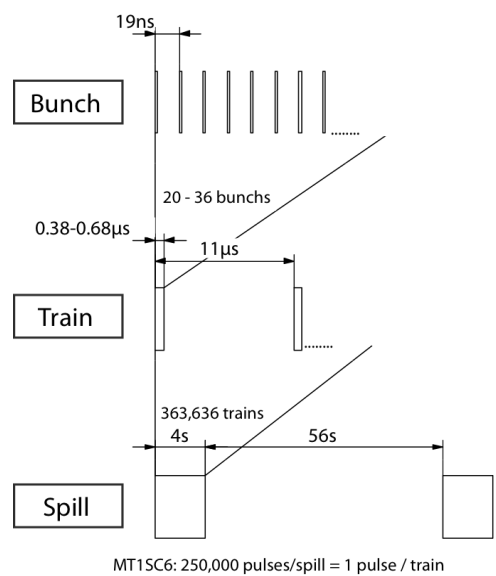

**Data Analysis**

First of all, charged particle events were separated by signal of a plastic scintillator in front of an NE213 scintillaor. The NE213 scintillator is sensitive to gamma-ray in addition to neutron. The neutron and gamma-ray events were separated by two gate integration method. Figure 3 shows an example of neutron and gamma-ray separation. Neutron events were 5 % of all events at 30° in the target-in measurement.

**Figure 3: Neutron and gamma-ray discrimination by an NE213 scintillator.**

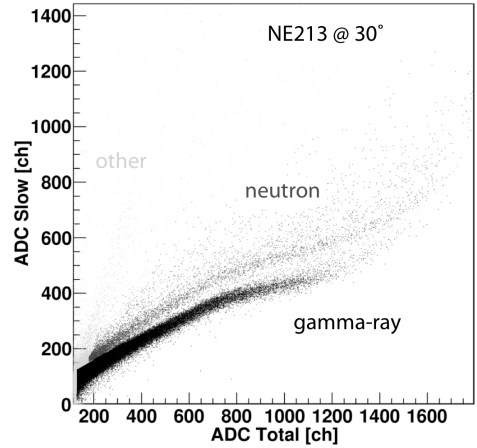

An example of time spectra is shown in Figure 4. One can see that several peaks with 400 ch for all charged particles, gamma-rays and neutrons. This interval shows the bunch of the beam time structure illustrated in Fig. 2.

**Figure 4: An example of time spectra of an NE213 scintillator.**

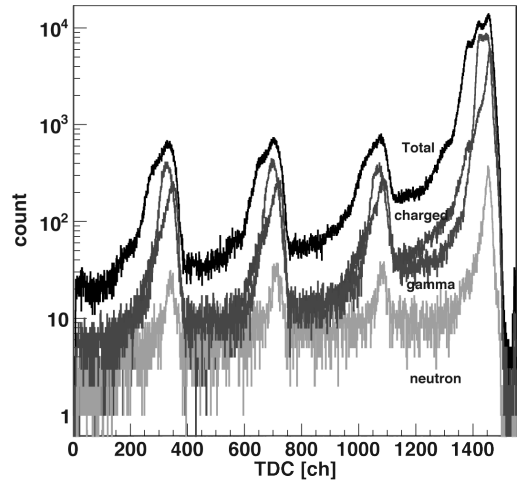

For separation of one proton incident events from multi-proton ones, the light output spectrum of plastic scintillators in front of the copper target was used. Figure 5 indicates the light output spectrum. The one proton incident events were distinguished from multi-proton ones. Events of 39 % were one proton incident events in the target-in measurement. Figure 6 shows the time spectra of one proton incident events (bm=1) after the separation process.

An example of light output spectra by one proton incident events separated by multi proton incident ones are seen in Figure 7. Neutron events were distinguished as seen in the figure.

**Figure 5: An example of ADC spectrum of the beam monitoring scintillator.**

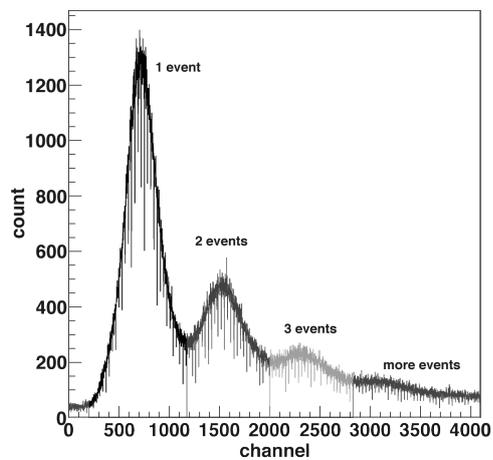

**Figure 6: An example of TDC spectra of an NE213 scintillator.**

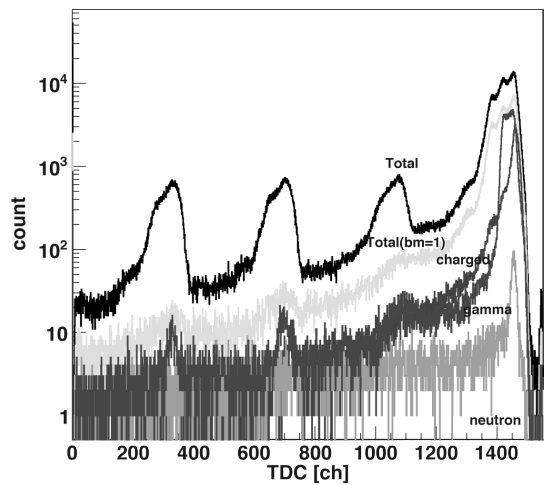

**Figure 7: An example of light output spectra of an NE213 scintillator.**

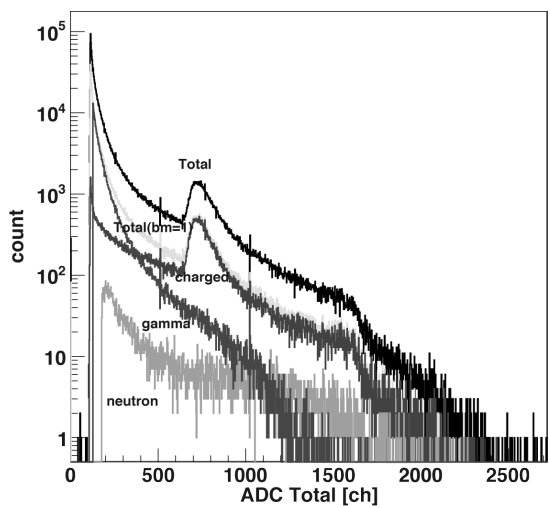

*Detection efficiency*

The neutron detection efficiency were usually calculated by the SCINFUL-QMD code [7] above 100 MeV of neutron energy. Hewever, the code dose not consider individual characteristics of each scintillator. In this experiment, the neutron detection efficiency for both scintillators were measured by continuous energy neutron beam at the WNR facility of Los Alamos Neutron Science Center (LANSCE) in Los Alamos National Laboratory in order to see the characteristics of both scintillators.

The WNR facility supplies pulsed neutron beam with spallation reaction of 800 MeV proton beam and a tungsten target below 700 MeV. Each neutron energy was determined by the time-of-flight technique between the spallation target and the NE213 scintillator. The incident neutron flux was obtained by the count of a $^{238}$U fission ionization chamber and the neutron induced fission cross section for $^{238}$U. Several types of neutron induced fission cross sections for $^{238}$U based on experimental data [8-10], calculation and evaluation [11-13] are presented in Figure 8. The neutron detection efficiency measured and derived by some neutron induced fission cross sections are shown in Figure 9 with calculated values by SCINFUL-QMD. The efficiency by SCINFUL-QMD overestimates the experimental ones between 60 and 200 MeV.

**Figure 8: Neutron induced fission cross sections for $^{238}$U.**

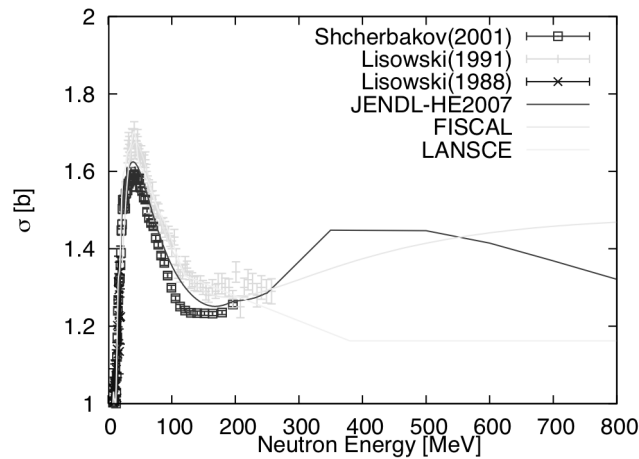

**Figure 9: Neutron detection efficiency of an NE213 scintillator at LANCE.**

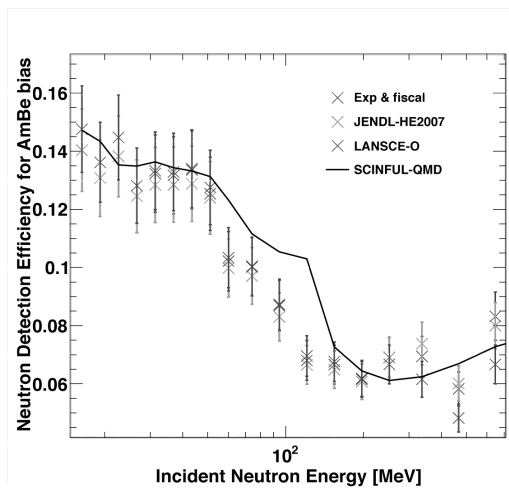

**Results**

The preliminary double-differential thick target neutron yields using the CAMAC DAQ system at 30˚ and 90˚ are shown in Figures 10 and 11, respectively with the calculated values by the PHITS [14] and the FULKA [15] codes. Both codes reproduce the shape of energy spectra. The experimental data at 30˚ were analyzed several kinds of neutron detection efficiencies. The PHITS code underestimates experimental neutron yields above 40 MeV in the direction. On the other hand, Both PHITS and FULKA codes slightly overestimate experimental data below 300 MeV at 90˚. The highest neutron energies were about 2 GeV at 30˚ and about 800 MeV at 90˚, respectively.

**Figure 10: 120 GeV proton induced neutron thick target yield at 30˚ using a CAMAC DAQ system.**

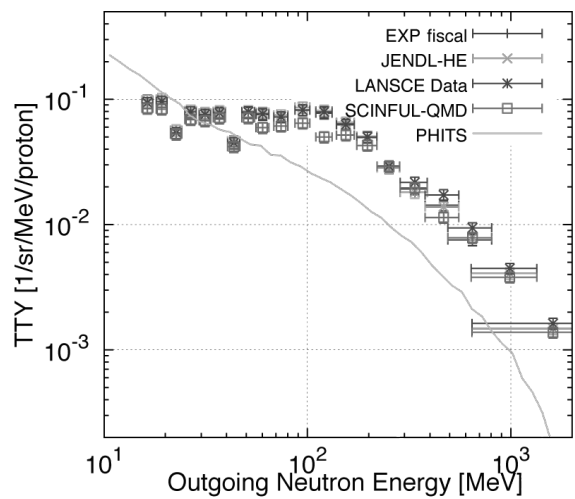

**Figure 11: 120 GeV proton induced neutron thick target yield at 90˚ using a CAMAC DAQ system.**

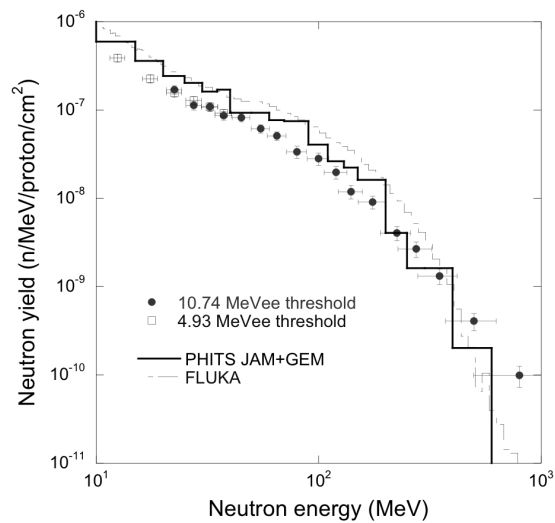

Proton and neutron energy spectra were retrieved by analyzing data obtained with the wave form digitizer DAQ. Figure 12 indicates the preliminary proton and neutron energy spectra emitted from the copper target with PHITS calculation results. Experimental data of neutron energy spectra by the wave form digitizer DAQ system is consistent with that by the CAMAC DAQ system. PHITS code generally reproduces the experimental proton up to 200 MeV and neutron data below 400 MeV, respectively.

**Figure 12: 120 GeV proton induced proton and neutron thick target yields at 90˚ using a wave form digitizer DAQ system.**

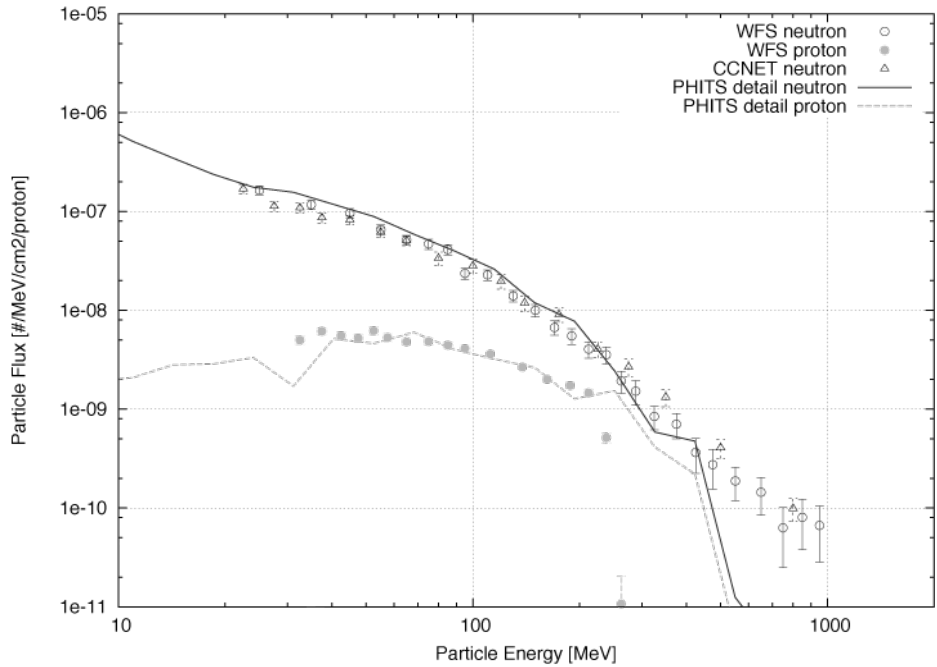

**Summary**

The thick target neutron yields and proton energy spectrum on a copper target for 120 GeV proton beam as a source term of high energy radiation behavior were feasibly measured. The experimental procedure using the ordinary CAMAC and the wave form digitizer DAQ system was confirmed. The apparatus would be applied to future measurements to obtain a systematic data set for secondary particle production on various target materials. The multi-particle transports code represent the shape of neutron and proton energy spectra. The calculation result by PHITS code underestimates the experimental data of thick target neutron yield at 30˚. The PHITS code generally reproduces proton and neutron energy spectrum at 90˚.

**Acknowledgement**

This work is supported by grant-aid of ministry of education (KAKENHI 21360473) in Japan. Fermilab is a U.S. Department of Energy Laboratory operated under Contract DE-AC02-07CH11359 by the Fermi Research Alliance, LLC.